\documentclass[runningheads]{llncs}
\usepackage{booktabs}
\usepackage{makecell}
\usepackage{multirow}
\usepackage{listings}
\usepackage{xcolor}
\usepackage{graphicx}
\usepackage{subcaption}
\usepackage{pgfplots}
\usepackage{framed}
\usepackage{amsmath}
\pgfplotsset{compat=newest}

\newcommand{\ie}{{\textit{i.e.}}, }
\newcommand{\eg}{{\textit{e.g.}}, }

\definecolor{codegreen}{rgb}{0,0.6,0}
\definecolor{codegray}{rgb}{0.5,0.5,0.5}
\definecolor{codepurple}{rgb}{0.58,0,0.82}
\definecolor{backcolour}{rgb}{0.95,0.95,0.92}

\lstdefinestyle{mystyle}{
  commentstyle=\color{codegreen},
  keywordstyle=\color{magenta},
  numberstyle=\tiny\color{codegray},
  stringstyle=\color{codepurple},
  basicstyle=\footnotesize,
  breakatwhitespace=false,         
  breaklines=true,                 
  captionpos=b,                    
  keepspaces=true,                 
  numbers=left,                    
  numbersep=5pt,                  
  showspaces=false,                
  showstringspaces=false,
  showtabs=false,                  
  tabsize=2,
  frame = single
}

\begin{document}
\title{SEED: Semantic Graph Based Deep Detection for Type-4 Clone}

\author{Zhipeng Xue \and
Zhijie Jiang\thanks{corresponding author} \and
Chenlin Huang\protect\footnotemark[1] \and
Rulin Xu \and
Xiangbing Huang \and
Liumin Hu
}
\authorrunning{X. Zhipeng et al.}

\institute{National University of Defense
Technology, China
\email{\{xuezhipeng19,jiangzhijie,clhuang,xurulin11\}@nudt.edu.cn\\hxbbing@hotmail.com,921406764@qq.com}}

\maketitle             
\begin{abstract}
Type-4 clones refer to a pair of code snippets with similar semantics but written in different syntax, which challenges the existing code clone detection techniques. 
Previous studies, however, highly rely on syntactic structures and textual tokens, which cannot precisely represent the semantic information of code and might introduce non-negligible noise into the detection models.
To overcome these limitations, we design a novel semantic graph-based deep detection approach, called SEED. For a pair of code snippets, SEED constructs a semantic graph of each code snippet based on intermediate representation to represent the code semantic more precisely compared to the representations based on lexical and syntactic analysis. To accommodate the characteristics of Type-4 clones, a semantic graph is constructed focusing on the operators and API calls instead of all tokens. Then, SEED generates the feature vectors by using the graph match network and performs clone detection based on the similarity among the vectors. 
Extensive experiments show that our approach significantly outperforms two baseline approaches over two public datasets and one customized dataset. Especially, SEED outperforms other baseline methods by an average of 25.2\% in the form of F1-Score. Our experiments demonstrate that SEED can reach state-of-the-art and be useful for Type-4 clone detection in practice.

\keywords{intelligent software engineering \and clone detection\and semantic graph \and graph neural network.}
\end{abstract}

\section{Introduction}

Code clones, widely existing in software systems (\eg 15\%-25\% in Linux kernel \cite{antoniol2002analyzing}), exert a significant impact on software maintenance and evolution (\eg fault localization \cite{zhang2017theoretical,li2006cp} and code refactor \cite{pizzolotto2020blanker,mazinanian2016jdeodorant}).
Typically, code clones can be categorized into four types \cite{roy2009comparison} based on different levels of similarity. \textbf{Type-4 clone} refers to syntactically dissimilar code snippets that implement the same semantic, which is the most challenging problem for traditional code clone detection techniques.

Currently, researchers have tried to consider Type-4 code clone detection as a classification task and solve it with deep learning methods \cite{white2016deep,zhang2019novel}. They build contextual embedding models of source code, form feature vectors for code representation, and then measure similarity among code vectors to detect code clones. For example, ASTNN \cite{zhang2019novel} presents a two-stage embedding approach based on recurrent neural network (RNN) to extract features from the abstract syntax tree (AST). TBCCD \cite{yu2019neural} likely links the AST with tokens to add more semantic information and generates the feature vector by tree-based LSTM. The performances of those approaches, however, are limited due to the following two main limitations. First, the existing studies rely heavily on syntactic structures (\eg AST) and cannot precisely represent the semantic information of code. Second, textual tokens commonly adopted for the code representation do not contain semantics that Type-4 clones require but introduce unnecessary noise data.

To explore an effective semantic-based solution for Type-4 clones, we propose a novel approach called \textbf{SEED} (\textbf{S}emantic-based cod\textbf{E} clon\textbf{E} \textbf{D}etector) in this paper. The key idea of SEED is to perform clone detection based on 1)~the code semantic structures rather than syntactic or lexical structures, and 2)~ emphasizing operator and API call tokens rather than universal tokens. First, SEED takes a code pair as input and constructs the semantic graph of each code based on intermediate representation (IR) \cite{zeng2021degraphcs,li2020automated} to represent the semantics of the source code. As the intermediary between high-level and assembly language, IR represents code as specific instructions and therefore is closer to the code semantics.
Then, SEED models the semantic graphs by using the graph match network (GMN) \cite{li2019graph} and generates feature vectors of source code.
Finally, SEED predicts Type-4 clone pairs based on the similarities among feature vectors. We evaluate the performance of our approach compared with two typical baseline approaches over two public datasets and one large-scale customized dataset. The results prove that SEED outperforms baseline approaches by over 25.2\% on average (reaching state-of-the-art) in the real-world scenario.

The main contributions of this paper are summarized as follows:
\begin{itemize}
	\item {we proposed a semantic-based deep learning approach SEED for Type-4 clone detection. SEED adopts the graph match network on the semantic graph, which is built from code semantic structures and enhanced by operator and API call tokens.}
	
	\item {We customized a Type-4 code clone dataset called CF-500\footnote{https://github.com/ZhipengXue97/SEED} to mitigate the threat posed by the lack of semantics of popular datasets (\i.e., POJ-104 and BigCloneBench). CF-500 consists of 500 functionalities, which is approximately 5 times the size of popular datasets.}
	
	\item {We evaluated the performance of our approach over two public datasets and 
	one customized dataset. The results indicate that SEED can achieve state-of-the-art performance and outperforms baseline approaches by over 25.2\% on average in a real-world scenario.}
\end{itemize}

The remainder of this paper is organized as follows. In Sec. \ref{related work} we survey the related work about code clone detection. In Sec. \ref{proposed method}, we illustrate the overview of SEED. In Sec. \ref{experiment}, we evaluate the performance of our approach with baselines by answering three research questions. In Sec. \ref{discuss}, we discuss the threats to the validity of the results. In Sec. \ref{conclusion}, we conclude our work.
\section{Related Work}
\label{related work}

As a critical problem in software maintenance, code clone detection has always been a hot spot for research. Traditional code clone detection approaches mainly focus on Type-1, 2, 3 code clones. They detect the code clone based on specific features such as tokens, metrics, and graphs. For example, sourcerer \cite{linstead2009sourcerer} performs clone detection based on the token. It obtains the code blocks with the least frequent tokens in code snippets, then indexes the code blocks and compares the blocks to find the clone pairs. Moreover, some methods use structure features such as AST, PDG, or CFG. Deckard \cite{jiang2007deckard} computes the feature vectors of ASTs and adopts the Locality Sensitive Hashing algorithm to detect cloned code. CCgraph \cite{zou2020ccgraph} converts the code snippets into PDGs, it then applies the Weisfeiler-Lehman kernel to compute the graph similarity and identifies the clone pairs. Although these methods use various information of the code, the significant information loss in feature generation leads to the limitation of their performance. Also, these methods rely heavily on syntax information, which makes them unable to handle Type-4 code clones. 

Since the deep learning perform impressive increment in natural language process, recently, researchers try to introduce deep learning in code clone detection. Different from traditional code clone detection approaches, the deep learning based code clone detection approaches convert the code feature into vector, and compare the similarity of the vectors. The deep learning based code clone detection approaches assign different weights to different parts of the code feature, which lead to a better performance on Type-4 code clone detection,  White et al. \cite{white2016deep} firstly introduces the deep learning method to code clone detection. They use a recurrent neural network to convert the textual information into the vector and learn the code representation from AST. Following this work, CDLH \cite{wei2017supervised} and TBCCD \cite{yu2019neural} search for deep learning models that are more suitable for the tree structure of the AST and propose their method which uses the tree-based deep learning model to handle AST. ASTNN \cite{zhang2019novel} proposes a novel two-step approach to represent the code snippet, using RNN to encode the AST of each statement first and transforming the AST encoding of all the statements into one vector to represent the code snippet. Oreo \cite{saini2018oreo}and Deepsim \cite{zhao2018deepsim} collect more than 20 features from code syntax, and transfer the collected features into vector. However, prior studies mainly relied on syntactic structure, as well as identifiers, leads to significant limitations in their performance on Type-4 clone detection. In contrast,  SEED focuses on the semantic structure of code and the operator and API call tokens.

\section{Proposed Method}
\label{proposed method}

\begin{figure*}[htbp]
	\centering
	\includegraphics[width=13cm]{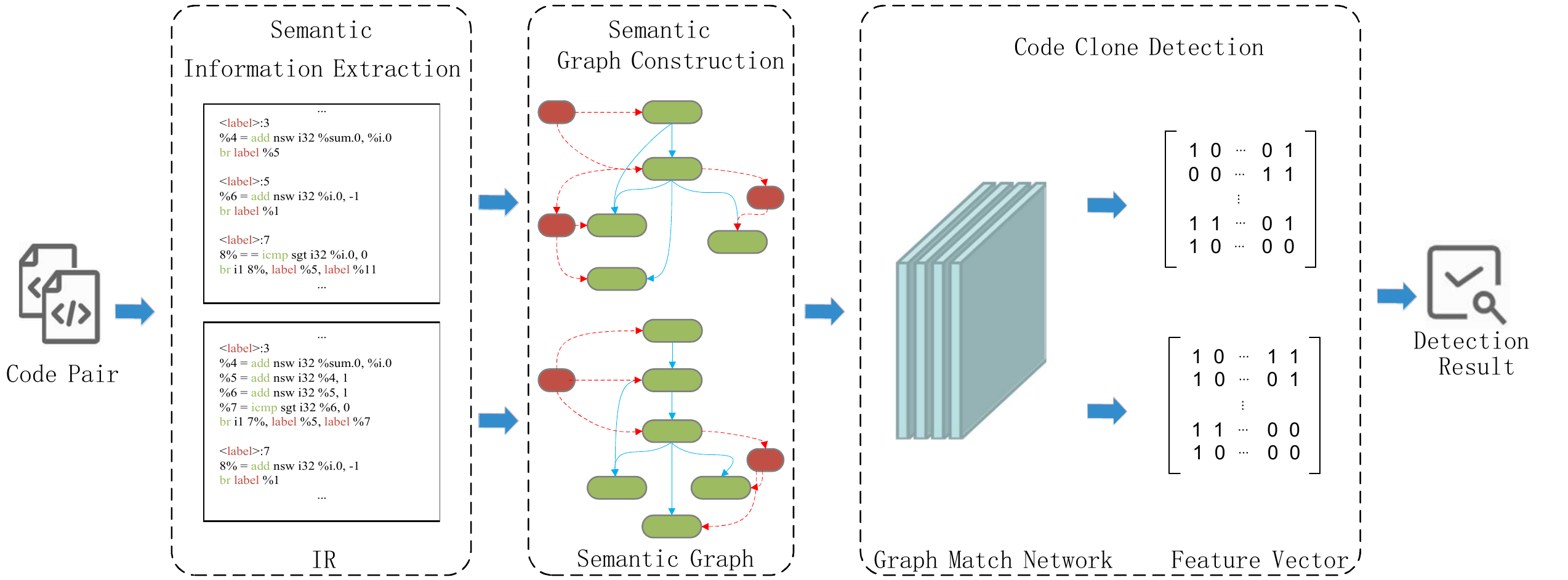}
	\caption{The Overview of SEED}
	\label{overview}
\end{figure*}

In this section, we propose a semantic-based deep graph learning method named SEED to cope with Type-4 clone detection. SEED detects Type-4 clones by constructing semantic graphs to represent code semantics while focusing on the operation semantics including operator and API call tokens.

\subsection{Overview}
Fig.~\ref{overview} shows the overview of SEED. SEED consists of three steps, semantic feature extraction, semantic graph construction, and code clone detection. During the first step, SEED takes a pair of code snippets as input and obtains the intermediate representation (IR) from the compiler. Then, in the second step, SEED constructs a semantic graph for each code snippet from IR to represent the code semantics. In the last step, SEED takes the semantic graphs as the input and transforms them into feature vectors using graph match network (GMN) \cite{li2019graph}. Subsequently, SEED predicts Type-4 clones based on whether the cosine similarity between the two feature vectors reaches a certain threshold.

\subsection{Semantic Feature Extraction}
\label{Semantic Feature Extraction}

SEED is designed to detect Type-4 clones based on code semantics. IR, as the language between the high-level programming language and the assembly language, converts the complex grammars of the code into basic instructions. Therefore, this version is closer to the developer's intention and can represent the code semantics more precisely.

In this paper, SEED supports \textit{C/C++} and \textit{Java} and extracts semantic features using \textit{LLVM}\footnote{https://llvm.org/}  and \textit{Soot}\footnote{http://soot-oss.github.io/soot/}. Since IR can only be generated from Compilable code, we use tools such as \textit{JCoffee} \footnote{https://github.com/piyush69/JCoffee} to complete uncompilable code and help them pass that obstacle.

\begin{figure*}[h]
  \centering
\includegraphics[width=13cm]{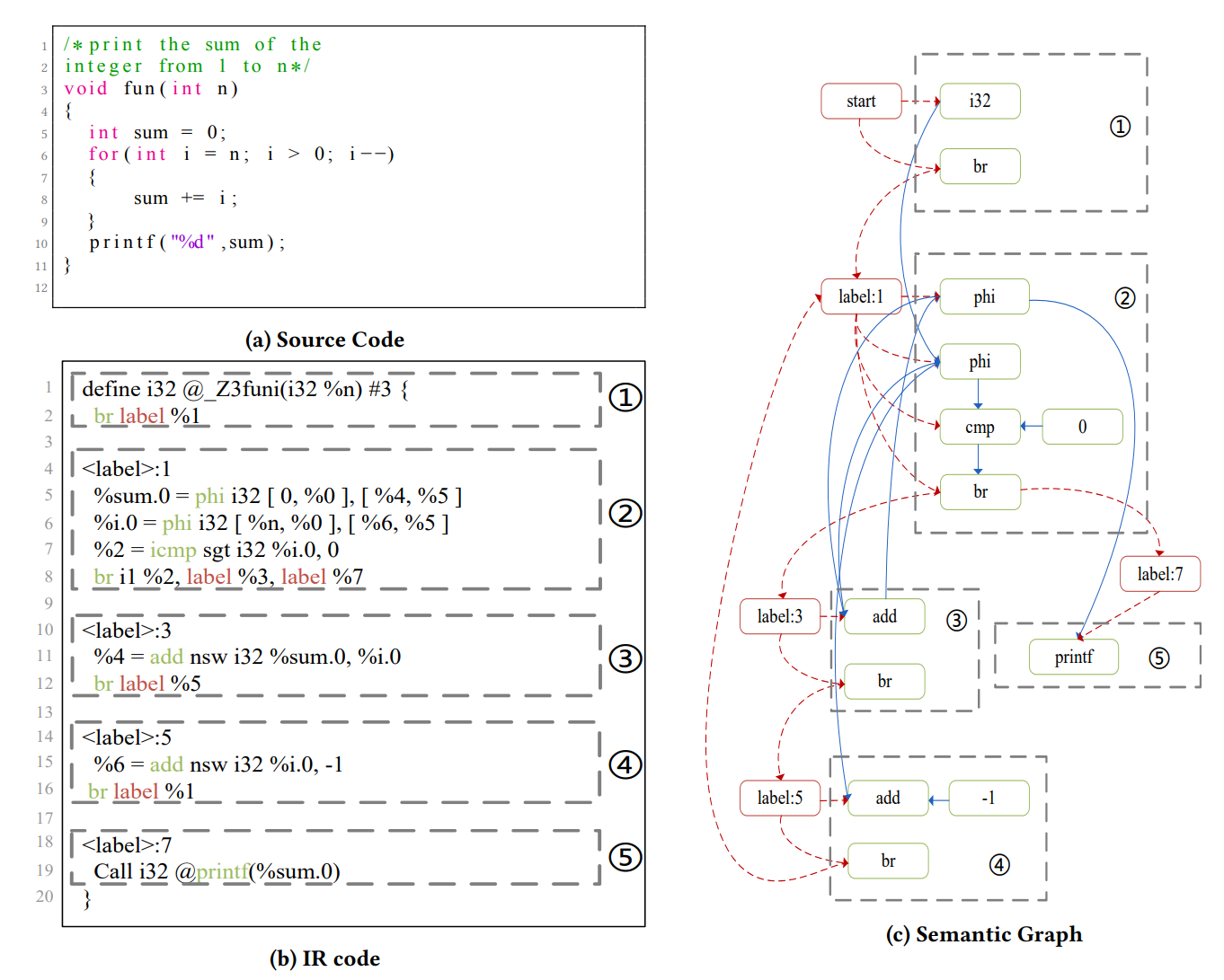}
\caption{an Illustrative Example of a Source Code Snippet and Its Corresponding IR Code and Semantic Graph}
\label{semantic graph example}
\end{figure*}

The operator refers to the symbol that tells the compiler or interpreter to perform specific mathematical, relational or logical operation. Most of the operator instructions in IR of \textit{C/C++} and \textit{Java} consist of three main parts: opcode, operand, and result, which means that the value of the operand is stored in the result after the operation of the opcode. Some instructions may miss the result or have more than two operands. Similarly, API call instructions in IR also consist of three main parts: API call, parameter, and result, which means that the API call output the result according to the input parameter. Following the execution order of the code snippet, IR divides instructions into several instruction blocks according to the branch instruction (i.e., {\tt br}) and uses a label to present the entry of each block. For example, Fig.~\ref{semantic graph example}(b) shows the IR for the source code in Fig.~\ref{semantic graph example}(a). Line 11 in Fig.~\ref{semantic graph example}(b) indicates that the sum of {\tt \%sum.0} and {\tt \%i.0} is stored in {\tt \%4}. Compared with token-based and AST-based Type-4 code clone detectors \cite{ben2018neural,linstead2009sourcerer}, instructions in IR focus on the operations on the variable, which describe the process of code execution and represent the intention of the developer (i.e., code semantics).

\subsection{Semantic Graph Construction}
\label{Semantic Information Representation}
To represent the code semantics, SEED combines the data flow and control flow to form the semantic graph based on IR while focusing on operator and API call tokens. In this section, we introduce semantic graph construction from two aspects: nodes and edges.

\subsubsection{Node}
\label{Node Extraction}
Different from previous studies \cite{white2016deep,wei2017supervised,yu2019neural}, which leverages all the textual tokens, the semantic graph constructed in this section only contains data type, operator, and API call tokens. We do not introduce identifier tokens in the semantic graph, since the identifier tokens in code are not reliable, the same identifier tokens can represent different semantics and lead to imprecise semantic representation in clone detection. In contrast, the same operator and API call tokens perform the same semantics in different codes.

Since each instruction contains only one operation (\ie one operator or one API call), we extract the operation from each instruction and take it as the node of the semantic graph. Since the label in IR represents the entry of each instruction block and contains the information of the control dependency of the code snippets, SEED also introduces the label as the node.
Moreover, to maintain the code semantics, SEED considers the constant and input data. If the operand of an instruction is a constant, we add a constant node next to its operation node. If the operand is input data, we add an input node with its data type. 

For example, as shown in Fig.~\ref{semantic graph example}(a) and Fig.~\ref{semantic graph example}(b), the line {\tt for (int i = n; i > 0; i$--$)} in the source code includes three components: {\tt int i=0}, {\tt i>0}, and {\tt i$--$}. Accordingly, the compiler splits the line into three different IR instructions in line 6, line 7 and line 15 of Fig.~\ref{semantic graph example}(b). These operations on variable {\tt i} include {\tt phi}, {\tt cmp}, and {\tt add}. SEED takes these three operations as three nodes. Similarly, the line {\tt printf("\%d", sum)} in Fig.~\ref{semantic graph example}(a) is recognized as a API call by compiler, and compiler generates corresponding IR instructions in line 19 of Fig.~\ref{semantic graph example}(b). SEED extracts the API call {\tt printf} and adds an operation node. For the constant {\tt 0} in the instruction of line 7, we add a constant node {\tt 0} next to the operation {\tt cmp}. Moreover, instructions in lines 4, 10, 14 and 18 represent the labels in IR, and SEED takes them as label nodes. The variable {\tt n} in the instruction of line 1 represents the input data; thus, SEED adds a node {\tt i32} to the semantic graph.

\subsubsection{Edge}
To integrate the data dependency and the control dependency, we add the data flow edge and the control flow edge to the semantic graph.

\textbf{Data Flow.}
For each instruction, if the result of it performs as the operand or parameter of another, we connect operation nodes of these two instructions with a data flow edge. Moreover, for each constant node or input node, we connect the data flow edges from these nodes to the operation nodes of their corresponding instructions.

For example, {\tt \%sum.0} in Fig.~\ref{semantic graph example}(b) is the result in line 5, while it is also the operand in line 7. Therefore, we add a data flow edge from node {\tt phi} in line 5 to node {\tt cmp} in line 7. Since {\tt \%sum.0} is also the parameter of API call {\tt printf}, we add a data flow edge from node {\tt phi} in line 5 to node {\tt printf} in line 19. Moreover, since {\tt 0} is the operand of operator {\tt cmp}, we add the data flow edge from the constant node {\tt 0} to the operator node {\tt cmp} in line 7.

\textbf{Control Flow.}
As discussed in Sec.~\ref{Semantic Feature Extraction}, IR divides instructions into several instruction blocks following the execution process of the code snippet. The instruction blocks can be used to represent the control dependencies of the code snippet. They are divided based on the branch instruction and use the label to represent the entry of each instruction block. The jumps between instruction blocks during the program execution are guided by operator {\tt br}. Therefore, we add the control flow edge from {\tt br} to corresponding label nodes. To illustrate the affiliation of instructions and instruction blocks, we also add control flow edges between operation nodes of instructions and label nodes of corresponding instruction blocks.

For example, in Fig.~\ref{semantic graph example}(b), since the operator {\tt add} and {\tt br} belong to the same instruction block as the label node {\tt label:3}, we connect the node {\tt label:3} to node {\tt add} and {\tt br} with control flow edges. Moreover, since node {\tt br} in line 8 is related to label nodes {\tt label:3} and {\tt label:7}, we add a control flow edge from node {\tt br} to label nodes {\tt label:3} and {\tt label:7}, respectively.

\subsection{Code Clone Detection}

SEED takes the clone detection task as a matching task. For a pair of code snippets, SEED generates the semantic graph of each code snippet into a feature vector and detects Type-4 clones based on the similarity between these two vectors. To adopt the graph structure of the semantic graph, we use the graph match network (GMN) \cite{li2019graph} to generate the feature vector of the semantic graph.

The input data of GMN are two semantic graphs $(G_1, G_2)$ of the code pair, each semantic graph $G=(V, E)$, where $V$ is the set of vertices and $E$ is the set of edges. First, we initialize the feature vector of each node as $h_i^{(0)}$ using the word2vec model \cite{mikolov2013efficient}. Moreover, we initialize data flow edges and control flow edges with different weights. Then, we calculate the node feature vectors over multiple iterations to learn the feature vectors representing the code semantics. 

For each iteration $t$, each node updates its feature vector $h_i^{(t)}$ based on its feature vector $h_i^{(t-1)}$ in iteration $t-1$, the message $m_i^{(t)}$ from the neighbor nodes in the same semantic graph, the similarity feature vector $\mu_i^{(t)}$ from another semantic graph. GMN uses a gated recurrent unit (GRU) \cite{cho2014learning} to update the feature vector in Eq.~\ref{hi}.

\begin{align}
\label{hi}
h_i^{(t)} &= \mathbf{GRU}(h_i^{t-1}, m_i^{(t)},\mu_i^{(t)})
\end{align}

For node $i$, $m_i^{(t)}$ refers to the message from its neighbors via the edges, allowing $h_i^{(t)}$ to obtain interrelationships between the node and the entire semantic graph.

\begin{align}
\label{neighbor}
m_i^{(t)}&=\sum_{j}\mathbf{SUM}({h}_{i}^{(t)}, {h}_{j}^{(t)}, {e}_{i j})
\end{align}

In Eq.~\ref{neighbor}, node $j$ is a neighbor of node $i$ in the same semantic graph, the message from node $j$ to node $i$ is calculated by weighted sum, and the weight is the feature vector of edge $e_{ij}$. Moreover, GMN adopts an attention mechanism to generate $h_i^{(t)}$ while referencing the semantic graph of another code snippet.  

\begin{align}
\label{attention_equation}
\alpha _{k\rightarrow i} &= \frac{exp(s_h(h_i^{(t-1)},h_k^{(t-1)}))}{  {\textstyle \sum_{k^{'}}^{}} exp(s_h(h_i^{(t-1)},h_{k^{'}}^{(t-1)})}\\
\mu_i^{(t)} &= \sum_{k}\alpha _{k\rightarrow i}(h_i^{(t-1)}-h_k^{(t-1)})
\end{align}

As shown in Eq.~\ref{attention_equation}, $s_h$ is a cosine similarity metric. $h_k^{(t-1)}$ represents the feature vectors of the nodes in another semantic graph. $\alpha _{k\rightarrow i}$ refers to the similarity between node $i$ and node $k$, which is used as the attention weight. $\mu_i^{(t)}$ aggregates the attention weights between the node $i$ and all the nodes in another semantic graph and represents the attention mechanism of node $i$. $\mu_i^{(t)}$ allows $h_i^{(t)}$ to represent the semantics of each node with a focus based on the difference between the two semantic graphs and consequently, helps the model represent the semantics of two code snippets more precisely. 

After $ T $ iterations, the feature vector of each node ${h}_{i}^{(T)}$ represents the semantics of each node and corresponding instructions. Subsequently, to represent the code semantics of the entire semantic graph, GMN aggregates the feature vector of each node into the feature vector of the semantic graph (${h}_{G}$) using a multilayer perceptron (MLP) \cite{li2015gated}. Furthermore, since the information of each node has a different contribution to the code semantics, GMN adopts the attention mechanism during the calculation of ${h}_{G}$.

  \begin{align}
  \label{mlp}
    \mathbf{h}_{G} &= \operatorname{MLP}_{G}\left(\sum_{i \in V} \sigma\left(\mathrm{MLP}_{\mathrm{gate}}\left(\mathbf{h}_{i}^{(T)}\right)\right) \odot \mathrm{MLP}\left(\mathbf{h}_{i}^{(T)}\right)\right)
    \end{align}
    
As shown in Eq.~\ref{mlp}, $\sigma(\mathrm{MLP}_{\mathrm{gate}}(\mathbf{h}_{i}^{(T)}))$ represents the attention mechanism, which assigns different weights to different node feature vectors. It is trained during the training process and guides the aggregation of node feature vectors. It generates the ${h}_{G}$ with focus, which enables it to represent code semantics more accurately.

Finally, we adopt cosine similarity to calculate the similarity between the feature vector of the two code snippets. By comparing the similarity of feature vectors and the threshold, we can predict whether the input code pair is a Type-4 clone. We choose the threshold empirically based on the validation set.

\section{Experiment}
\label{experiment}

In this section, we conduct experiments to evaluate our approach by answering the following research questions.\\
{\bfseries RQ1: How does SEED perform against baseline approaches? }\\
Previous studies achieve good experimental results on their datasets \cite{zhang2019novel,yu2019neural}. To compare the performance of SEED with those of previous studies, we followed the experimental setup of previous studies to test the model performance.\\
{\bfseries RQ2: How does SEED perform when implemented on a more diversified dataset? }\\
The existing datasets, although they contain numerous code pairs, the number of their semantics is relatively small. This makes code clone detectors only exposed to limited semantics, thereby posing a threat to the training and testing of the model. Accordingly, we constructed a more diversified dataset and tested the performance of SEED to understand if SEED can achieve a consistent result.\\
{\bfseries RQ3: How effective are the different semantic graph construction strategies of SEED at Type-4 clone detection?}\\
Since SEED constructed the semantic graph while focusing on the operation. To evaluate the performance of emphasizing operator and API call tokens in semantic graph construction and make sure the semantic graph construction strategy in SEED is the best one, we carried out an ablation study on the semantic graph construction method.

\subsection{Experiment Setup}
\subsubsection{Datasets}
\label{Experiment_Datasets}
In our experiment, we use two public datasets Big-CloneBench\cite{svajlenko2014towards} and POJ-104\cite{mou2016convolutional}, and one larger-scale customized dataset called CF-500. The overall information of datasets is listed in Table \ref{ALL Dataset}.

BigCloneBench is built by mining frequently used code semantics from \textit{Java} project dataset, IJAdataset-2.0. We select 11,799 compilable code snippets covering 43 semantics from the BigCloneBench as one of our datasets. POJ-104 \cite{mou2016convolutional} is a widely used dataset to evaluate the performance of Type-4 code clone detection. It is collected from an open judging platform POJ\footnote{http://poj.org/}, which contains many programming problems and corresponding submissions. Since submissions of the same problem in the open judging platform usually have the same semantic and different syntactic structures and therefore can be classified into Type-4 clone pairs. POJ-104 contains 104 problems and 500 submissions written in \textit{C} for each problem. Since the submissions will be compiled and executed by the open judging platform, all these submissions are compilable and thereby can be used as our dataset.

Although the existing datasets contain numerous clone pairs, they have relatively limited semantics, resulting in limited exposure to less semanticity and posing a threat to the training and testing of the model. To alleviate such a threat, we built a more diversified dataset named CF-500. We collected CF-500 from the open judging platform Codeforces\footnote{http://codeforces.com/}. CF-500 contains more than 23,000 code snippets written in \textit{C}, covering 500 problems.

Since each two code snippets can construct a code pair, the number of code pairs in a dataset can be higher than 10,000,000 if using all possible combinations. Due to the vast number of code pairs, we randomly downsample the code pairs to build our datasets. For all the training sets, we sample 100,000 code pairs randomly with the proportion of the clone pairs and nonclone pairs as 1:1. For the validation set and the test set, we randomly select 10,000 code pairs from the remaining pairs.

\subsubsection{Baseline Approaches}

We reproduced two state-of-the-art approaches, ASTNN and TBCCD, to compare with our approach. ASTNN \cite{zhang2019novel} proposes a novel two-step approach to represent the code snippet, using RNN to encode the AST of each statement first and transforming the AST encoding of all the statements into one vector to represent the code snippet. TBCCD \cite{yu2019neural} uses position-aware character embedding (PACE) technology to embed tokens. PACE takes the token embedding and AST as the input and generates the feature vector using tree-based LSTM to represent the code snippet. Both of them likely use the similarity of feature vectors to detect code clones. We do not compare with FA-AST \cite{wang2020detecting} or Deepsim \cite{zhao2018deepsim}, since they only support \textit{JAVA} and report similar results compared to TBCCD in their paper.

\begin{table}[htbp]
\centering
   \caption{\label{ALL Dataset}Overall Information for Datasets}  
    \begin{tabular}{cccc}
        \toprule    Datasets & Language & Semantics & \makecell[c]{Code \\ snippets} \\  
        \midrule   BigCloneBench & JAVA & 43 & 11,799 \\
        POJ-104 & C & 104 & 52,000\\ 
        CF-500 & C & 500 & 23,146\\
        \bottomrule   
    \end{tabular} 
 \end{table}

\subsection{Answer to RQ1: Overall Performance}

In this experiment, we compare SEED's performance against two baseline approaches. Previous studies \cite{wang2020detecting,wei2017supervised}  simply split code pairs into the training, validation, and test sets to proceed with the experiment. Although such an approach ensures that code pairs in the three sets are not the same, it does not consider the semantics of the code snippet. In other words, different implementations of the same semantics may be split into different sets, making it possible for the model to see the code snippets under the same semantics in the test set during the training process. However, the semantics vary in real-world software, and it is impossible to include all semantics in the training set. To better evaluate the performance of SEED, we split the training, validation, and test sets from different code semantics.

Following the setting in TBCCD \cite{yu2019neural}, we use two public datasets, BigCloneBench and POJ-104, to evaluate the model performance. We construct the training set and the validation set from their first 15 problems and 10 problems. Instead of using all the remaining problems as the test set, we divided them into 6 and 3 test sets to evaluate the robustness of SEED when testing different semantics. We illustrate the experimental results in Table~\ref{RQ1}. Columns \textit{P}, \textit{R}, and \textit{F1} represent precision, recall, and F1-Score, respectively.

We find that SEED significantly outperforms both baseline approaches on all test sets. On POJ-104, SEED obtains an average F1-Score of 0.62, which is higher than the F1-Score of ASTNN (0.45) and TBCCD (0.50). Compared with TBCCD, SEED outperforms by at least 17.5\% (from 0.40 to 0.47) in the form of the F1-Score in the test set with problem IDs 16-30 and even by approximately 30\% (from 0.53 to 0.68) in the form of F1-Score in the test set with problem IDs 61-75. For BigCloneBench, similarly, SEED achieves an average F1-Score of 0.54, which significantly outperforms that of baseline approaches. In particular, in terms of precision, SEED outperforms baselines by 74.4\% on average. In the real-world software repositories, the accuracy of clone pairs reported by SEED is higher than the accuracy of baselines. Compared with the performance on POJ-104, the F1-Scores of baselines in BigCloneBench drop by approximately 25\%, while those of SEED drop by less than 15\% because BigCloneBench is collected from a practical software environment in which the identifiers and API calls vary. Although exciting improvement of SEED in precision, recall, and F1-score, we must acknowledge that SEED performs poorer robustness than baselines, since the F1-score of SEED drops 34\% (from 0.71 to 0.47), while that of TBCCD and ASTNN only drops 28\% (from 0.56 to 0.40) and 29\% (from 0.51 to 0.36) in different test sets.

\begin{table*}[htbp]  
\centering
   \caption{\label{RQ1}Result of SEED and Other Baselines on the POJ-104 and BigCloneBench Dataset}  
    \begin{tabular}{c|c|ccccccccc}  
     \toprule 
     \multirow{2}*{Datasets} & \multirow{2}*{\makecell{Problem IDs\\ for Testing}}& \multicolumn{3}{c}{ASTNN} & \multicolumn{3}{c}{TBCCD} & \multicolumn{3}{c}{SEED}\\
       \cline{3-5} \cline{6-8} \cline{9-11}
       & & P & R & F1 & P & R & F1 & P & R & F1\\
       \hline
       \multicolumn{1}{c|}{\multirow{6}*{POJ-104}}&16 $-$ 30 & 0.32 & 0.42 & 0.36 & 0.35 & 0.46 & 0.40 & \textbf{0.41} & \textbf{0.57} & \textbf{0.47} \\
       \multicolumn{1}{c|}{}&31 $-$ 45 & 0.44 & 0.61 & 0.51 & 0.51 & 0.62 & 0.56 & \textbf{0.74} & \textbf{0.68} & \textbf{0.71} \\
       \multicolumn{1}{c|}{}&45 $-$ 60 & 0.41 & \textbf{0.60} & 0.49 & 0.59 & 0.48 & 0.53 & \textbf{0.71} & \textbf{0.60} & \textbf{0.64} \\
       \multicolumn{1}{c|}{}&61 $-$ 75 & 0.41 & 0.65 & 0.51 & 0.58 & 0.49 & 0.53 & \textbf{0.78} & \textbf{0.69} & \textbf{0.68} \\
       \multicolumn{1}{c|}{}&76 $-$ 90 & 0.38 & \textbf{0.63} & 0.47 & 0.61 & 0.44 & 0.51 & \textbf{0.71} & 0.57 & \textbf{0.63} \\
       \multicolumn{1}{c|}{}&(91 $-$ 104) + 16 & 0.37 & 0.45 & 0.40 & 0.48 & 0.45 & 0.46 & \textbf{0.61} & \textbf{0.55} & \textbf{0.58} \\
      \multicolumn{1}{c|}{}&average & 0.39 & 0.56 & 0.45 & 0.52 & 0.49 & 0.50 & \textbf{0.66} & \textbf{0.61} & \textbf{0.64} \\
       \hline
       \multicolumn{1}{c|}{\multirow{3}*{BigCloneBench}}&12 $-$ 22 & 0.28 & 0.39 & 0.32 & 0.20 & \textbf{0.92} & 0.33 & \textbf{0.58} & 0.44 & \textbf{0.50} \\
       \multicolumn{1}{c|}{}&23 $-$ 33 & 0.34 & 0.40 & 0.37 & 0.27 & \textbf{0.91} & 0.41 & \textbf{0.72} & 0.57 & \textbf{0.63} \\
       \multicolumn{1}{c|}{}&34 $-$ 44 & 0.25 & 0.30 & 0.27 & 0.17 & \textbf{0.67} & 0.27 & \textbf{0.44} & 0.51 & \textbf{0.47} \\
      \multicolumn{1}{c|}{}& average & 0.29 & 0.36 & 0.33 & 0.21 & \textbf{0.83} & 0.38 & \textbf{0.58} & 0.51 & \textbf{0.54} \\
       \bottomrule
    \end{tabular} 
 \end{table*}

\subsection{Answer to RQ2: Larger-scale Experiment}

In this experiment, we aimed to alleviate the threat caused by the small number of the semantics of the existing dataset and evaluate the performance of our approach over the more diversified datasets, CF-500. Since the experiment setting in previous studies \cite{yu2019neural,wei2017supervised} and RQ1 only use a limited size of datasets, we also use the entire POJ-104 and BigCloneBench dataset to do an experiment and analyze the result.

To understand the robustness of SEED in different sizes of dataset, We keep the validation set and test set unchanged and train the model on varied training sets. For BigCloneBench, we extract a validation set from semantic IDs 32-37 and a test set from semantic IDs 38-44. The training sets are built from semantics IDs 2-11, 2-21, and 2-31. For POJ-104, we set problem IDs 76-90 as the validation set and problem IDs 91-104 as the test set. Then, we create several training sets of different sizes. The training sets cover problem IDs 1-15, 1-30, 1-45, 1-60, and 1-75. For CF-500, we set problem IDs 401-450 and 451-500 as the validation set and test set, and problem IDs 1-100, 1-200, 1-300, and 1-400 as training sets, providing us with a series of training sets of increasing size from 10 to 400.

\begin{table*}[htbp]  
\centering
  \caption{\label{RQ2}Result of SEED and Other Baselines on the More Diversified Datasets}  
   \begin{tabular}{c|c|ccccccccc}  
    \toprule 
        \multirow{2}*{Datasets} & \multirow{2}*{\makecell{Problem IDs\\ for Training}}& \multicolumn{3}{c}{ASTNN} & \multicolumn{3}{c}{TBCCD} & \multicolumn{3}{c}{SEED}\\
      \cline{3-5} \cline{6-8} \cline{9-11}
      & & P & R & F1 & P & R & F1 & P & R & F1\\
      \hline
      \multicolumn{1}{c|}{\multirow{5}*{POJ-104}}&1 $-$ 15 & 0.36 & 0.48 & 0.41 & 0.40 & 0.63 & 0.49 & \textbf{0.51} & \textbf{0.64} & \textbf{0.57} \\
      \multicolumn{1}{c|}{}&1 $-$ 30 & 0.44 & 0.52 & 0.50 & 0.51 & \textbf{0.71} & 0.59 & \textbf{0.74} & 0.66 & \textbf{0.70} \\
      \multicolumn{1}{c|}{}&1 $-$ 45 & 0.49 & 0.69 & 0.57 & 0.61 & \textbf{0.71} & 0.65 & \textbf{0.80} & 0.67 & \textbf{0.73} \\
      \multicolumn{1}{c|}{}&1 $-$ 60 & 0.55 & \textbf{0.80} & 0.65 & 0.69 & 0.66 & 0.67 & \textbf{0.78}& 0.74 & \textbf{0.76} \\
      \multicolumn{1}{c|}{}&1 $-$ 75 & 0.61 & \textbf{0.83} & 0.66 & 0.70 & 0.69 & 0.70 & \textbf{0.77} & 0.78 & \textbf{0.78} \\
      \hline
      \multicolumn{1}{c|}{\multirow{3}*{BigCloneBench}}&2 $-$ 11 & 0.30 & 0.33 & 0.31 & 0.21 & \textbf{0.71} & 0.32 & \textbf{0.50} & 0.55 & \textbf{0.52} \\
      \multicolumn{1}{c|}{}&2 $-$ 21 & 0.34 & 0.47 & 0.39 & 0.30 & \textbf{0.69} & 0.42 & \textbf{0.57} & 0.68 & \textbf{0.62} \\
      \multicolumn{1}{c|}{}&2 $-$ 31 & 0.36 & 0.55 & 0.43 & 0.33 & \textbf{0.69} & 0.45 & \textbf{0.68} & 0.68 & \textbf{0.68} \\
      \hline
      \multicolumn{1}{c|}{\multirow{4}*{CF-500}}&1 $-$ 100 & 0.63 & \textbf{0.85} & 0.72 & 0.71 & 0.77 & 0.74 & \textbf{0.82} & 0.74 & \textbf{0.78} \\
      \multicolumn{1}{c|}{}&1 $-$ 200 & 0.62 & \textbf{0.87} & 0.73 & 0.73 & 0.76 & 0.74 & \textbf{0.80} & 0.80 & \textbf{0.80} \\
      \multicolumn{1}{c|}{}&1 $-$ 300 & 0.62 & \textbf{0.89} & 0.74 & 0.73 & 0.77 & 0.75 & \textbf{0.81} & 0.82 & \textbf{0.81} \\
      \multicolumn{1}{c|}{}&1 $-$ 400 & 0.61 & \textbf{0.94} & 0.74 & 0.74 & 0.76 & 0.75 & \textbf{0.81} & 0.83 & \textbf{0.82} \\
      \bottomrule
   \end{tabular} 
\end{table*}

As illustrated in Table \ref{RQ2}, we find that SEED outperforms baseline approaches in all datasets. In BigCloneBench and POJ-104, SEED achieves a significant performance improvement by 54\% and 14\% on average in terms of the F1-Score, respectively. In particular, despite the greater number of semantics in CF-500, SEED still outperforms other approaches by 8\% in the form of the F1-Score. Similar to the result of RQ1, SEED achieves the highest precision in baseline approaches while maintaining a similar recall, validating the result we discussed in RQ1 and proving that our approach can achieve a constant result when tested on more diversified datasets.

Furthermore, the results also prove that our extended dataset can alleviate the threat posed by the low number of the semantics of datasets. In BigCloneBench, with the increasing size of the dataset, the performance of SEED increases by up to 31\%, 24\%, and 36\% in the form of F1-Score, Recall, and Precision, respectively. Likely, on POJ-104, with the increasing size of the dataset, the metrics of SEED increase by up to 37\%, 21\%, and 51\%, respectively. In contrast, with the increasing size of the dataset in CF-500, the precision of SEED remains stable, while the F1-Score and recall increase 5\% and 24\%, respectively. To understand such improvement, we analyze code pairs that are correctly determined only after the dataset is expanded. The code snippets in BigCloneBench are collected from the real-world project and implement the semantics using mainly the API calls. The clone detection model can better understand the semantics of API calls by training on a more semantic-diverse dataset. For POJ-104 and CF-500, code snippets focus on the algorithm. Training on a more semantically diverse dataset enables the model to better identify the core structures of the code snippets and ignore irrelevant ones. Therefore, the result indicates that training code clone models with a semantics diverse dataset can help improve the model performance.

To establish a better understanding of the effect of dataset size on the performance of the model, we compared the performance of SEED with baseline approaches using datasets of different sizes. We found that the performance of all models increases as the size of the training set grows. Meanwhile, as the size of the dataset increases, the speed of performance improvement gradually decreases and peaks when the dataset reaches a certain size, indicating that the performance improvement caused by the size of the dataset is limited when it reaches a certain number. Moreover, we find that SEED outperforms baseline approaches over various sizes of datasets and can achieve the same performance as baselines by training on a smaller dataset, also verifying the validity of our approach.

\subsection{Answer to RQ3: Ablation Study}

An ablation study usually refers to comparing the performance of different strategies. Since we focus on the operator and API call tokens when constructing the semantic graph, to verify the validity of this method, we carried out an ablation study to evaluate the semantic graph construction method.
We use the same dataset as RQ2. For the experimental group, we constructed different semantic graphs by combining the identifiers and data types in the semantic graphs. The setting of these semantic graphs is as follows:
\begin{itemize}
  \item SEED+identifier: SEED+identifier uses not only the operator and API call tokens, but also identifier tokens of variables as the nodes in the semantic graph and connects data flow edges from each identifier node to its corresponding operation node. 
  \item SEED+type: SEED+type still introduces variables as nodes in the semantic graph, while it replaces the identifier tokens of variables by the data type tokens of them.
  \item SEED: SEED removes all the variable or data type token, and only uses the operator and API call tokens as the node in the semantic graph.
\end{itemize}

The results of our experiments are shown in Table \ref{RQ3}. To explore the reason for different performances when using different semantic graph construction methods, we collected the changed sizes of the constructed semantic graph and compared the differences among them. The changed sizes of different semantic graphs are shown in Table \ref{RQ3_comp}. The column \textit{ratio} refers to the maximum decrease ratio between SEED and the other two models.

First, we compared the performance of SEED+identifier and SEED+type to discuss the influence of using tokens without instructive semantics in Type-4 Clone.

\begin{table*}[htbp]  
\centering
   \caption{\label{RQ3}Result of SEED and Other Settings of Semantic Graph on the Different Datasets}  
    \begin{tabular}{c|c|ccccccccc}  
     \toprule 
     \multirow{2}*{Datasets} & \multirow{2}*{\makecell{Problem IDs\\ for Training}}& \multicolumn{3}{c}{SEED+identifier} & \multicolumn{3}{c}{SEED+type} & \multicolumn{3}{c}{SEED}\\
       \cline{3-5} \cline{6-8} \cline{9-11}
       & & P & R & F1 & P & R & F1 & P & R & F1\\
       \hline
       \multicolumn{1}{c|}{\multirow{1}*{POJ-104}}&1 $-$ 75 & 0.77 & 0.71 & 0.74 & \textbf{0.78} & 0.75 & 0.76 & 0.77 & \textbf{0.78} & \textbf{0.78} \\
       \hline
       \multicolumn{1}{c|}{\multirow{1}*{BigCloneBench}}&2 $-$ 31 & 0.50 & 0.57 & 0.53 & 0.64 & 0.61 & 0.62 & \textbf{0.68} & \textbf{0.68} & \textbf{0.68} \\
       \hline
       \multicolumn{1}{c|}{\multirow{1}*{CF-500}}&1 $-$ 400 & 0.74 & \textbf{0.83} & 0.79 & 0.77 & 0.82 & 0.80 & \textbf{0.81} & \textbf{0.83} & \textbf{0.82} \\
       \bottomrule
    \end{tabular} 
 \end{table*}
The results show that using identifiers leads to a slight performance drop of approximately 2.6\% over POJ-104 and CF-500. However, using identifiers reduces the model performance in the form of F1-Score by approximately 17\% in BigCloneBench. To understand such differences, we studied the semantic graph size of the three models. From Table \ref{RQ3_comp}, we found that after replacing identifiers by their data types (SEED+identifier and SEED+type), the size of vocabulary decreased 58,941.
This result indicates that although the code snippet in BigCloneBench has good naming rules such as the \textit{Camel case}
, it still introduces noise into the dataset and leads to the out-of-vocabulary problem. In contrast, code snippets in POJ-104 and CF-500 are written by only one programmer for one simple task, making the programmer often use simple characters such as {\tt i}, {\tt j}, etc., resulting in a small size of the vocabulary and alleviating the out-of-vocabulary problem and preventing the model from a significant performance drop.

Second, the performance between SEED+type and SEED illustrates the effectiveness of the size of the semantic graph. For POJ-104 and CF-500, the F1-Score increases by approximately 2\%, while for BigCloneBench, the F1-Score increases by 10\%. We studied the reason for this difference. Table \ref{RQ3_comp} shows the content of the semantic graph in SEED+type and the content of the semantic graph in SEED.
We found that by constructing a semantic graph without identifiers except for constants, the number of operand nodes and data flow edges in the semantic graph decrease by over 60\% in all three datasets because SEED uses the only operation to construct the semantic graph, which significantly reduces the size of the semantic graph. With a smaller size semantic graph, the GMN model can focus on learning the semantic information (i.e., operation) in the semantic graph, resulting in a more accurate feature vector generated by GMN and consequently the improvement of the model performance.

In conclusion, the results in RQ3 indicate that the semantic graph construction strategy in SEED outperforms alternative strategies in terms of most metrics over all three datasets, proving that the semantic graph focusing on the operations can better represent the code semantics.

 \begin{table}[htbp]  
 \centering
  \caption{\label{RQ3_comp}The Changed Size of SEED's Semantic Graph and Other Settings of Semantic Graph on the Different Datasets}  \small
  \begin{tabular}{c|c|cccc}  
    \toprule 
    Datasets & Characteristic & \makecell{SEED+\\identifier} & \makecell{SEED+\\type} & SEED & Ratio\\
      \hline
      \multicolumn{1}{c|}{\multirow{3}*{POJ-104}}& vocabulary size& 6204 & 2350 & 1981 & 0.68\\
      \multicolumn{1}{c|}{}& operand node & 126.83 & 126.83 & 43.84 & 0.65\\
      \multicolumn{1}{c|}{}& dataflow edge&426.48 & 426.48 & 153.10 & 0.64\\
      \hline
      \multicolumn{1}{c|}{\multirow{3}*{\makecell{BigClone-\\Bench}}}& vocabulary size& 87,534 & 28,593 & 13,666 & 0.83\\
      \multicolumn{1}{c|}{}& operand node & 97.44 & 97.44 & 38.13 & 0.61\\
      \multicolumn{1}{c|}{}& dataflow edge & 145.93 & 145.93 & 51.58 & 0.64\\
      \hline
      \multicolumn{1}{c|}{\multirow{3}*{\makecell{CF-500}}}& vocabulary size& 7392 & 3175 & 2022 & 0.72\\
      \multicolumn{1}{c|}{}& operand node & 74.51 & 74.51 & 24.77 & 0.67\\
      \multicolumn{1}{c|}{}& dataflow edge&294.74 & 294.74& 97.98 & 0.67\\
      \bottomrule
  \end{tabular} 
\end{table}

\section{Threats to Validity}
\label{discuss}

In this section, we discuss the threat to the validity of our approach. SEED may suffer from two threats to validity as follows: 

\textbf{Internal Validity.} During semantic graph construction, we emphasize operator and API call tokens of IR instruction, based on the assumption that the semantic of operator and API call tokens perform more robust compared to identifier tokens. Different API calls may have similar functionality or in reverse, which affects the performance of SEED. To verify the validity of the assumption and the strategy in semantic graph construction, we do an ablation study in Sec.\ref{experiment}.

\textbf{External Validity.} We built a Type-4 code clone dataset CF-500 in our experiment. To make sure different problems refer to different functionalities, we compare the description of each problem and discard problems with the same descriptions. However, we can not guarantee whether different problem descriptions mean different functionalities.

\section{Conclusion}
\label{conclusion}

In this study, we presented a semantic-based deep detection approach, SEED, to detect Type-4 code clones. SEED focused on the semantic structure of code and the operator and API call tokens. To alleviate the threat posed by the small number of functionalities in the previous dataset, we constructed a dataset, CF-500, containing 23,146 code implementations of 500 functionalities. This dataset is nearly five times the size of the existing dataset. Extensive experiments over two public datasets and one customized dataset show that our approach, compared to the 3 baseline approaches, achieves state-of-the-art performance. Our approach can be applied in practice to assist with software maintenance.

In the future, we aim to further improve the performance of the tool supporting your approach. To better represent the semantics of the source code, we can filter out more less-semantic content from the semantic graph, and merge more semantic features in it. Another potential extension to our work is to leverage other GNN models, which can better compare the semantic graphs and embed them into feature vectors.
\section{Acknowledgements}
The authors would like to thank the anonymous reviewers for
their insightful comments. This work was substantially supported
by National Natural Science Foundation of China (No. 61872373 and 61872375).

\bibliographystyle{splncs04}
\bibliography{references}

\end{document}